# Structural properties and phase diagram of the La($Fe_{1-x}Ru_x$)AsO system

A. Martinelli[1,*], A. Palenzona[1], I. Pallecchi[1], C. Ferdeghini,[1] M. Putti,[1,2] S. Sanna[3], C. Curfs[4], C. Ritter[5]

[1] *SPIN-CNR, C.so Perrone 24,I- 16152 Genova – Italy*

[2] *Department of Physics, Università di Genova, Via Dodecaneso 33, I-16146 Genova, Italy*

[3] *Department of Physics "A. Volta," Università di Pavia-CNISM, I-27100 Pavia, Italy*

[4] *European Synchrotron Radiation Facility, 6 rue Jules Horowitz, F-38043 Grenoble, France*

[5] *Institute Laue - Langevin, 6 rue Jules Horowitz, F-38042 Grenoble Cedex 9, France*

**Abstract**

Structural refinement, lattice micro-strain and spontaneous strain analyses have been carried out on selected members of the La($Fe_{1-x}Ru_x$)AsO system using high resolution neutron and synchrotron powder diffraction data. The obtained results indicate that the character of the tetragonal to orthorhombic structural transition changes from first order for $x$ = 0.10, possibly to tricritical for $x$ = 0.20, up to second order for $x$ = 0.30; for $x \geq 0.40$ symmetry breaking is suppressed, even though a notable increase of the lattice micro-strain develops at low temperature. By combining structural findings with previous muon spin rotation data, a phase diagram of the La($Fe_{1-x}Ru_x$)AsO system has been drawn. Long-range ordered magnetism occurs within the orthorhombic phase ($x \leq 0.30$), whereas short-range magnetism appears to be confined within the lattice strained region of the tetragonal phase up to $x < 0.60$. The direct comparison between the magnetic and the structural properties indicates that the magnetic transition is always associated to the structural symmetry breaking, although confined to a local scale at high Ru contents.

## 1. Introduction

The class of materials referred to as Fe-based superconductors attracted much interest in the last years thanks to the discovery of a relatively high superconducting transition temperature ($T_c$) in LaFeAsO after electron doping induced by F-substitution.[1] Several families of new compounds were soon discovered, characterized by different structures and stoichiometries; in particular compounds characterized by the general formula *RE*FeAsO (*RE*: rare earth) crystallize at room temperature in the tetragonal system (*P*4/*nmm* space group) and undergo, however, on cooling a structural transition at $T_s$, producing an orthorhombic structure (*Cmme* space group).[2,3,4,5,6,7] The occurrence of the orthorhombic phase is accompanied by magnetic ordering of the Fe sub-lattice in undoped compounds.

[*] Corresponding author: alberto.martinelli@spin.cnr.it

First structural investigations on F-substituted samples claimed that the progressive electron-doping suppresses the structural symmetry breaking and the magnetic ordering as well.[3,4,6] Conversely a recent accurate synchrotron powder diffraction analysis ascertained that the structural effect of F-substitution is limited to a reduction of the orthorhombic distortion, with the symmetry breaking still being active even at optimal doping;[8] a latest NMR investigation confirmed such a result.[9]

Magnetic frustration,[10,11] nematic correlations[12],[13] and orbital ordering[14,15,16,17] are the most credited mechanisms proposed to explain the occurrence of symmetry breaking. Theoretical investigations revealed that the anisotropic properties observed above the structural transition can be originated by orbital ordering, irrespective of whether long-range magnetic order is present or not, and then the concept of nematic orbital order has been introduced.[18] A recent micro-structural analysis pointed out that a micro-strain along the *hh*0 direction takes place on cooling in the tetragonal phase of both doped and un-doped *RE*FeAsO compounds, increasing on cooling down to $T_s$.[19] Just above the structural transition micro-strain reaches its maximum value and then is abruptly suppressed by symmetry breaking; this behaviour is compatible with a scenario foreseeing orbital ordering as the driving parameter of the symmetry breaking, where nematic behaviour is induced by the tendency towards ordering of Fe 3*d* orbitals.[19]

Ruthenium is iso-electronic with iron and hence its progressive dilution in *RE*FeAsO compounds can yield useful hints into the comprehension of their structural and physical properties, as well as their interplay. Ru substitution in *RE*FeAsO compounds progressively suppresses magnetism without inducing superconductivity.[20,21] X-ray-absorption spectroscopy analyses evidenced that local disorder induced by the Ru substitution is mainly confined within the FeAs layer, whereas the *RE*O layer sustains a relative order.[22]

Theoretical calculations revealed that in the $RE(Fe_{1-x}Ru_x)AsO$ systems Ru atoms do not sustain any magnetic moment and Ru substitution progressively frustrates Fe moments.[23] Experimental investigations on optimally electron-doped $Sm(Fe_{1-x}Ru_x)As(O_{0.85}F_{0.15})$ samples showed that in the Fe-rich region superconducting and normal-state properties are strongly affected by disorder induced by Ru substitution; in the Ru-rich one the system is metallic and strongly compensated and the presence of Ru frustrates the magnetic moment on Fe ions.[23] A subsequent investigation revealed a re-entrant static magnetic order which degrades the superconducting ground state; it was then concluded that in this system the two order parameters compete, producing a nanoscopic phase separation.[24]

A recent investigation on the Pr(Fe,Ru)AsO system ascertained that Ru substitution suppresses the structural and magnetic transition characterizing PrFeAsO; in addition a negative thermal expansion was observed and was related to the absence of superconductivity in these compounds.[25] In the same system it has been suggested a complete suppression of magnetism for $Pr(Fe_{0.33}Ru_{0.67})AsO$.[20] In the Ce(Fe,Ru)AsO system AFM ordering at the transition metal sub-lattice is observed up to $x \leq 0.6$, whereas a further increase of Ru content induces a Pauli paramagnetic behaviour.[26]
The same samples of the $La(Fe_{1-x}Ru_x)AsO$ series described in the present work were previously characterized by several analytical techniques. The magneto-transport properties as well as the calculated band structure are reported in ref. [27,28]. Muon spin rotation (μSR) and NMR measurements evidenced that Ru substitution causes a progressive reduction of the magnetic transition temperature, of the magnetic order parameter, without leading to the onset of superconductivity, and yields a progressive decrease of the density of states at the Fermi level.[29] In particular this system behaves as a spin-diluted system with competing exchange interactions.
In this paper we report a detailed structural characterization of $La(Fe_{1-x}Ru_x)AsO$ compounds, obtained by high-resolution synchrotron and neutron powder diffraction data. As a conclusion the phase diagram of the La(Fe,Ru)AsO system has been drawn, combining structural and μSR data.

**2. Experimental**

Poly-crystalline $La(Fe_{1-x}Ru_x)AsO$ ($0.00 \leq x \leq 0.80$) samples were prepared reacting pre-synthesized LaAs with high purity $Fe_2O_3$, $RuO_2$, Fe, Ru in a similar way as described in ref. [23].
In order to check the [Fe]/[Ru] ratio, amounts of the samples were embedded in cold-setting epoxy resin, underwent metallographic preparation and were analyzed by energy dispersive X-ray spectroscopy (EDS; OXFORD X-Max$^N$). As a result the actual Ru content resulted only a very few percent lower than the nominal composition.
Preliminary phase identification was performed by X-ray powder diffraction at room temperature (XRPD; PHILIPS PW3020; Bragg-Brentano geometry; $CuK_{\alpha}$; range 15 – 120° $2\theta$; step 0.020° $2\theta$; sampling time 10 s). Neutron powder diffraction (NPD) analysis was carried out at the Institute Laue Langevin in Grenoble, using the high-resolution D2B diffractometer ($\lambda$ = 1.596 Å); NPD patterns were collected on selected samples ($x$ = 0.10, 0.20, 0.30, 0.40) between 10 and 150 K. Neutron thermo-diffractograms were collected using the high-intensity D1B diffractometer ($\lambda$ = 2.524 Å) on $x$ = 0.10, 0.20, 0.30 samples around the structural and magnetic transition temperatures. Synchrotron powder diffraction (SPD) analysis was carried out on selected samples ($x$ = 0.10, 0.30, 0.40, 0.50, 0.80) at the ID31 beam-line ($\lambda$ = 0.35443 Å) of the European Synchrotron Radiation Facility (ESRF) in Grenoble. In this case, in order to carefully evaluate the temperature at which the structural transition takes place, high statistic thermo-diffractograms of the 110 diffraction peak (tetragonal indexing) were also acquired on heating in a continuous scanning mode for samples with $x$ = 0.10, 0.30, 0.40; in fact the 110 tetragonal peak splits into the 020 + 200 orthorhombic lines on cooling, signing the symmetry breaking.

Structural refinement was carried out according to the Rietveld method[30] using the program FULLPROF;[31] refinements were carried out using a file describing the instrumental resolution function. In the final cycle the following parameters were refined: the scale factor; the zero point of detector; the background (parameters of the 5$^{th}$ order polynomial function); the unit cell parameters; the atomic site coordinates not constrained by symmetry; the atomic displacement parameters; the anisotropic strain parameters.

From the refined parameters micro-structural properties and spontaneous strain can be inspected. The micro-structure was investigated by using the refined anisotropic strain parameters and analyzing the broadening of diffraction lines by means of the Williamson-Hall plot method.[32] Generally, in the case where size effects are negligible and the micro-strain is isotropic, a straight line passing through all the points in the plot and through the origin has to be observed, where the slope provides the micro-strain: the higher the slope the higher the micro-strain. If the broadening is not isotropic, size and strain effects along particular crystallographic directions can be obtained by considering different orders of the same reflection.

## 3. Results

### *3.1 Analysis of the SPD thermo-diffractograms*

The upper panel of Figure 1, on the left, shows the high statistic thermo-diffractograms of the 110 diffraction peak (tetragonal indexing) of the La($Fe_{0.90}Ru_{0.10}$)AsO sample and, on the right, the thermal evolution of its broadening (measured as Lorentzian strain broadening). A marked increase of the peak broadening is observed below ~ 160 K, but an exact evaluation of the structural transition temperature is prevented. In fact at $T > T_s$ lattice micro-strain along $hh0$ progressively broadens the tetragonal 110 line as the structural transition is approached (see § 3.3). With the occurrence of the symmetry breaking, unresolved 110 peak splitting takes place for a low amplitude of the orthorhombic distortion. The crossover from the a micro-strained regime to an unresolved peak splitting then prevents an accurate determination of $T_s$, that, on the other hand, can be reliably evaluated by analyzing the spontaneous strain (see § 3.4).

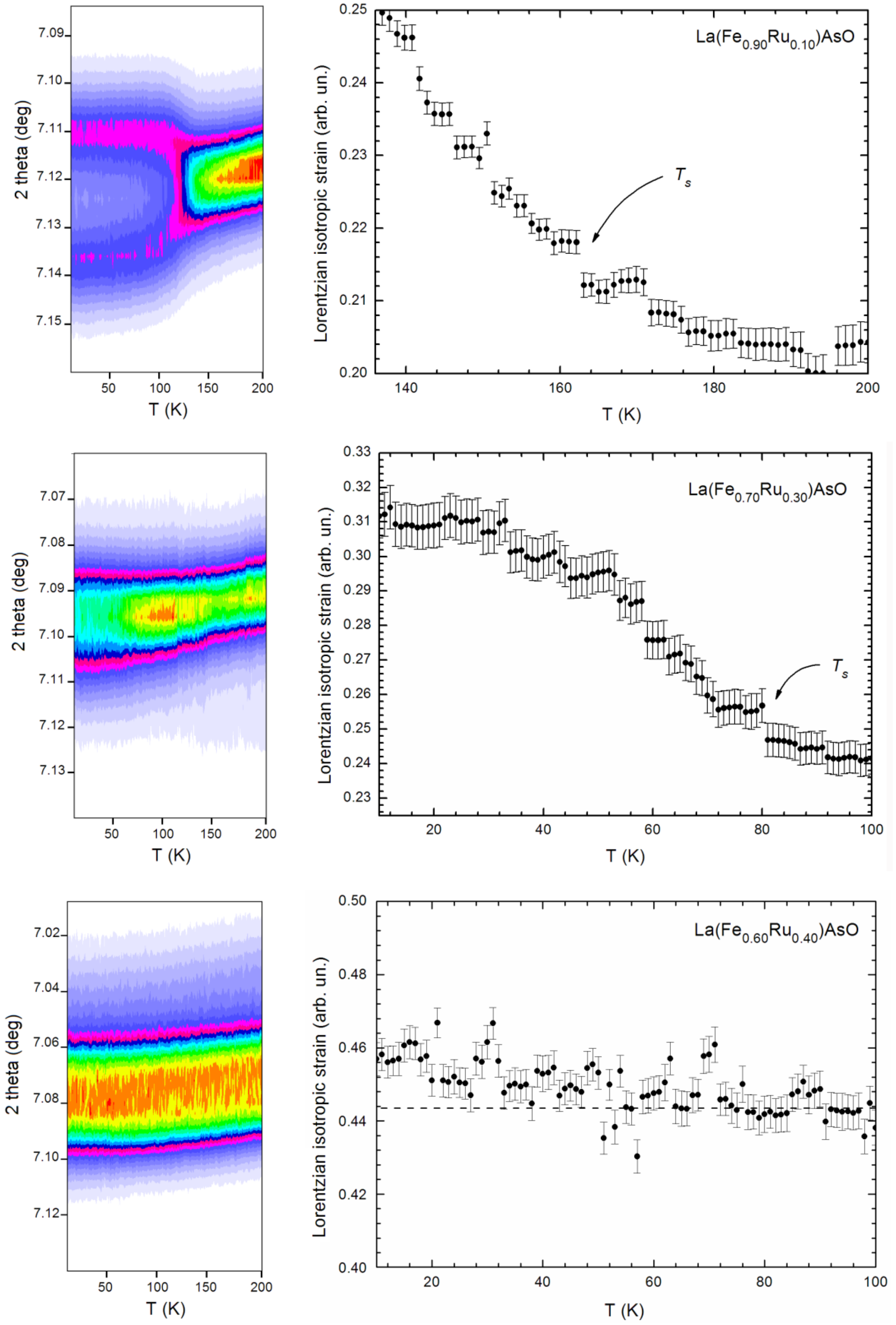

Figure 1: On the left: Thermo-diffractograms of the tetragonal 110 diffraction peak (SPD data) for samples with $x$ = 0.10, 0.30 and 0.40. On the right: thermal evolution of the corresponding Lorentzian strain broadening; in the lower panel the dotted line is the linear fit of higher temperature data for the $La(Fe_{0.60}Ru_{0.40})AsO$ sample, evidencing the continuous faint broadening of the peak below ~ 80 K.

With the increase of the Ru content, peak splitting is progressively suppressed on account of the decrease of the orthorhombic distortion amplitude. In any case an abrupt increase of the peak broadening is evident in the $La(Fe_{0.70}Ru_{0.30})AsO$ sample at $T_s$ ~ 80 K (Figure 1, middle panel), indicating that the structural transition is still taking place. Also in this case the exact determination of $T_s$ can be obtained by the thermal evolution of the spontaneous strain (see § 3.4).

For the $La(Fe_{0.60}Ru_{0.40})AsO$ sample no evident abrupt change can be appreciated in the whole inspected thermal range (Figure 1, lower panel); as a consequence it can be concluded that in this sample the structural transition is completely suppressed. A slight increase of the strain broadening can be detected below ~ 80 K; the significance of this broadening and its origin is deferred to § 3.3, where a more reliable micro-structural analysis is presented, as obtained by using the whole NPD pattern data.

These results suggest that the structural transition is progressively hindered by Ru substitution, down to its complete suppression for $0.30 < x < 0.40$; moreover these findings are in agreement with those of Yiu *et al.*,[25] that locate the structural transition at 75 K for $x = 0.33$ and report its complete suppression for $x = 0.40$ in the homologous $Pr(Fe_{1-x}Ru_x)AsO$ system.

### *3.2 Full pattern analysis and structural refinement*

Structural data at 300 and 10 K as obtained from the refinement of SPD data are reported in Tables 1 and 2, respectively, whereas Figure 2 shows the Rietveld refinement plot of the $La(Fe_{0.70}Ru_{0.30})AsO$ sample as obtained from SPD data collected at 300 K, selected as representative.

Table 1: Structural parameters of $La(Fe_{1-x}Ru_x)AsO$ samples, as refined from SPD data collected at 300 K; space group *P*4/*nmm*, origin choice 2; La and As atoms at 2*c* site, Fe and Ru atoms at 2*b* site, O atoms at 2*a* site.

| | $x = 0.10$ | $x = 0.30$ | $x = 0.40$ | $x = 0.50$ | $x = 0.80$ |
|---|---|---|---|---|---|
| *a* (Å) | 4.0414(1) | 4.0562(1) | 4.0658(1) | 4.0751(1) | 4.1023(1) |
| *c* (Å) | 8.7409(1) | 8.6924(1) | 8.6621(1) | 8.6347(1) | 8.5562(2) |
| *z* La | 0.1415(1) | 0.1414(1) | 0.1413(1) | 0.1410(1) | 0.1412(1) |
| *z* As | 0.6512(1) | 0.6514(1) | 0.6518(1) | 0.6517(2) | 0.6522(2) |
| $R_{Bragg}$ (%) | 5.74 | 5.46 | 5.28 | 5.91 | 7.60 |

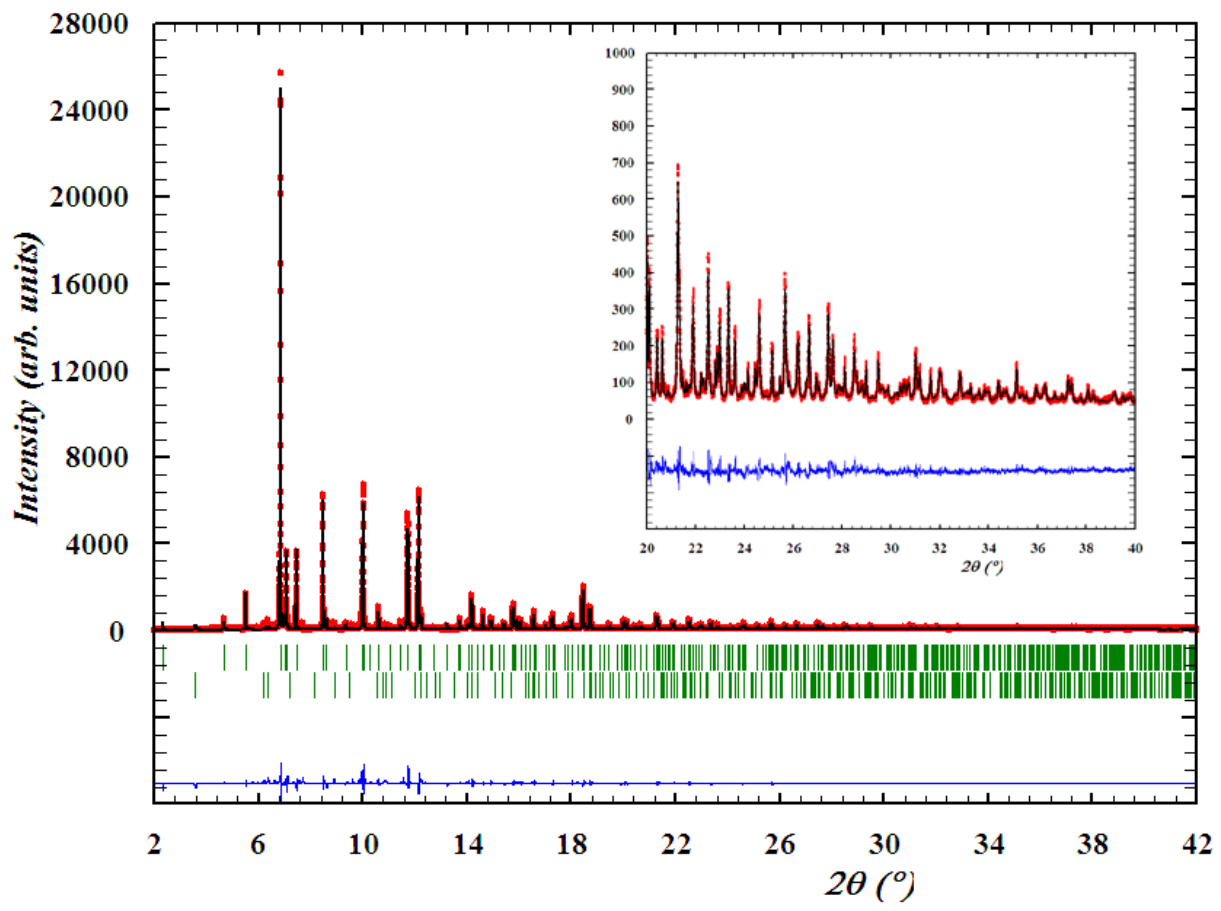

Figure 2: Rietveld refinement plot of $La(Fe_{0.70}Ru_{0.30})AsO$ (SPD collected at 300 K); the inset shows an enlarged view of the plot in the high angular region of the pattern.

A closer inspection of the diffraction patterns reveals the presence of a small amount of $La(OH)_3$ in the samples, that were taken into account during refinements. Both cell edges follows the Vegard's rule, but in opposite ways: the lattice parameter *a* linearly increases, according to the relationship $a = 0.088x + 4.0312$, as the average ionic radius at the transition metal site increases by Ru substitution. Conversely the lattice parameter *c* linearly decreases, according to the relationship $c = -0.266x + 8.7689$, a phenomenon that can be ascribed to the decrease of the As-*TM*-As bond angle (*TM*: transition metal) with the increase of Ru content.[23] These linear changes indicate that the substitution at the *TM* site is characterized by a tendency towards structural relaxation, that is the inter-atomic distances tend to be the similar for both Fe and Ru atoms.[33] This is in agreement with X-ray-absorption spectroscopy analyses carried out on $Sm(Fe_{1-x}Ru_x)As(O_{0.85}F_{0.15})$ samples, that measured a difference between Fe-As and Ru-As bond lengths of ~ 0.03 Å only, half than expected.[22]

Table 2: Structural parameters of $La(Fe_{1-x}Ru_x)AsO$ samples, as refined from SPD data collected at 10 K. Samples with $x = 0.10$ and 0.30 crystallize in the *Cmme* space group; La and As atoms at 4*g* site, Fe and Ru atoms at 4*b* site, O at atoms at 4*a* site. Samples with $x = 0.40$, 0.50 and 0.80 crystallize in the *P*4/*nmm* space group, origin choice 2; La and As atoms at 2*c* site, Fe and Ru atoms at 2*b* site, O atoms at 2*a* site.

| | ***x* = 0.10** | ***x* = 0.30** | ***x* = 0.40** | ***x* = 0.50** | ***x* = 0.80** |
|---|---|---|---|---|---|
| *a* (Å) | 5.7160(1) | 5.7301(1) | 4.0592(1) | 4.0693(1) | 4.0972(1) |
| *b* (Å) | 5.6942(1) | 5.7227(1) | / | / | / |
| *c* (Å) | 8.7138(1) | 8.6693(1) | 8.6399(1) | 8.6150(2) | 8.5375(1) |
| *z* La | 0.1421(1) | 0.1418(1) | 0.1417(1) | 0.1415(1) | 0.1413(1) |
| *z* As | 0.6508(2) | 0.6510(1) | 0.6514(2) | 0.6515(3) | 0.6521(4) |
| R-Bragg (%) | 4.27 | 4.41 | 6.50 | 8.49 | 10.80 |

Figure 3 shows the evolution of the cell edges between 10 and 300 K in samples with $x$ = 0.10, 0.30 and 0.40. Whatever the composition, the $c$ axis homogeneously decreases on cooling and then tends to a constant value at lower temperatures; this behaviour differs from what reported for the homologous Pr(Fe,Ru)AsO system, where an unusual negative thermal expansion of the $c$ axis was observed below ~ 60 K and related to the suppression of superconductivity.[25] This different thermal dependence of the lattice parameters can be explained by the different magnetic nature of the $La^{3+}$ and $Pr^{3+}$ ions. In fact in PrFeAsO a magnetic transition occurs below 11 K, where the $Pr^{3+}$ ions magnetically order, but an interplay between Fe and Pr moments is already active below 40 K.[34] This conclusion is corroborated by the fact that a negative thermal expansion is also reported for NpFeAsO, characterized by an antiferromagnetic structure of the Np sub-lattice.[35]

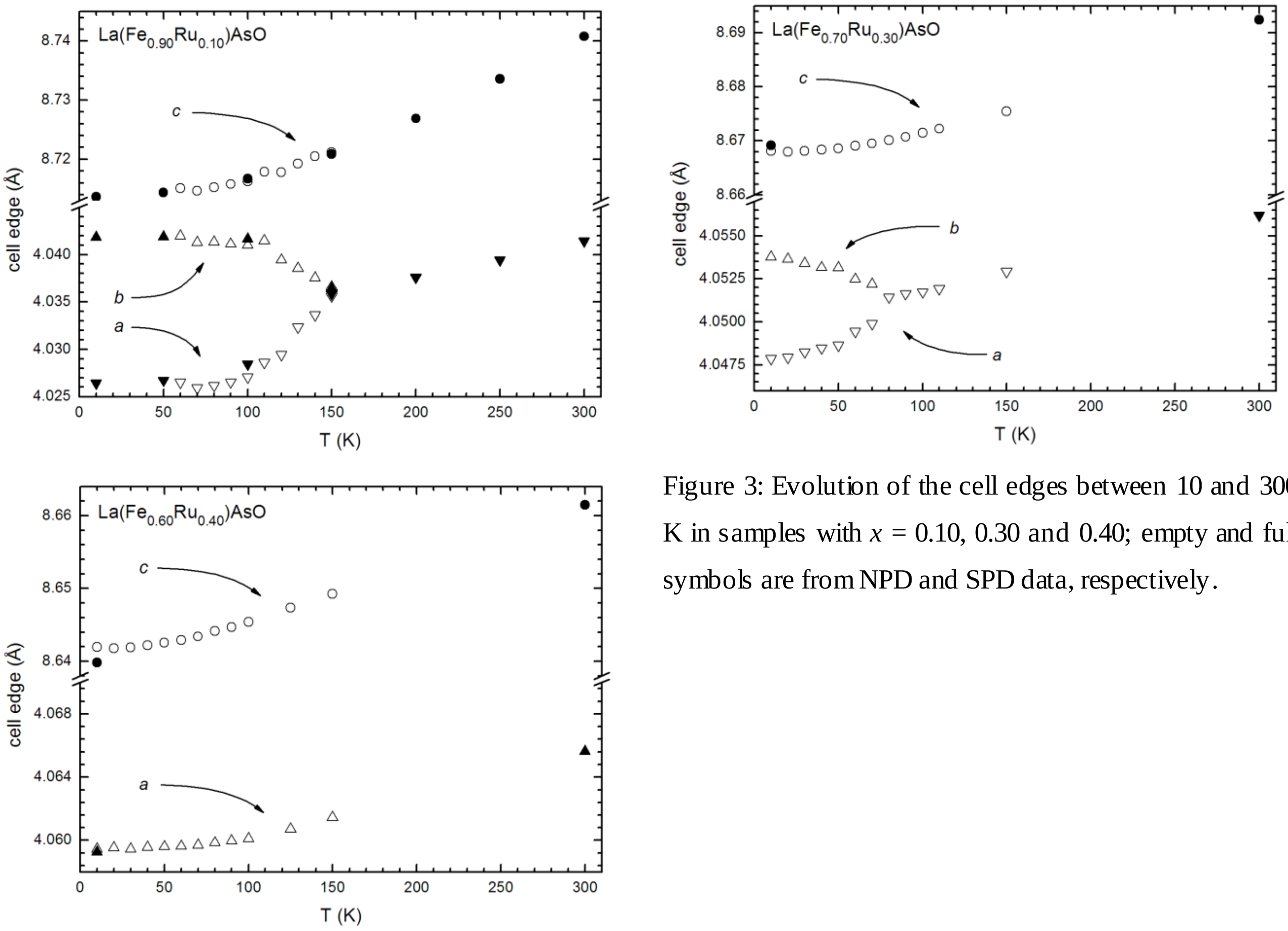

Figure 3: Evolution of the cell edges between 10 and 300 K in samples with $x$ = 0.10, 0.30 and 0.40; empty and full symbols are from NPD and SPD data, respectively.

In order to investigate in detail the thermal expansion behaviour in the La(Fe,Ru)AsO system, the cell volume of selected samples has been fitted between 10 and 300 K, using a Grüneisen second-order approximation for the zero-pressure equation of state:[36]

(1)

Where and ; is a dimensionless Grüneisen parameter of the order of unity; is the compressibility and its derivative with respect to applied pressure; is the zero temperature limit of the unit cell volume; is the internal energy calculated by the Debye approximation:

(2)

Where N is the number of atoms in the unit cell; is the Boltzmann's constant; is the Debye temperature. The fitting was carried out assuming $= 1.03 \cdot 10^{11}$ Pa, which is the experimental bulk modulus value extracted from high pressure SPD measurements on $SmFeAs(O_{0.93}F_{0.07})$,[37] and leaving , and as free parameters. Figure 4 shows the resulting fitting curves for three samples; in all cases, values turn out to be in the range 230 K – 250 K, in fair agreement with the value calculated in LaFeAsO ( = 282 K) from heat capacity data.[38] It can be seen that the Grüneisen law well accounts for the observed temperature dependence of the volume in the whole temperature range explored by the experiment and no departure is detected at low temperature, at odds with the findings reported for the Pr(Fe,Ru)AsO system, where an unusual negative thermal expansion is observed below ~ 60 K.[25]

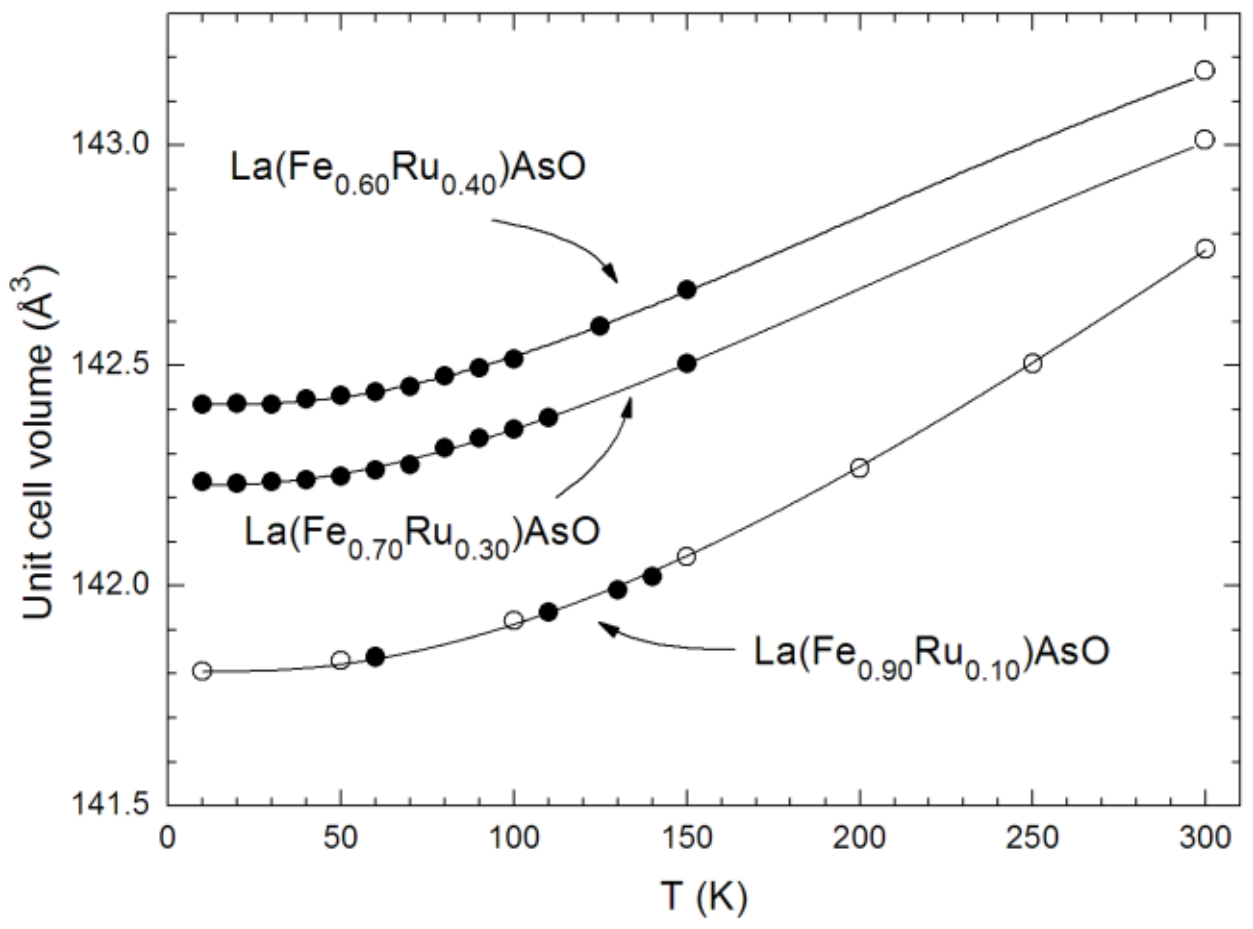

Figure 4: Unit cell volume as a function of temperature in selected $La(Fe_{1-x}Ru_x)AsO$ compounds; the solid line shows the best fit to a second-order Grüneisen approximation (open symbols: SPD data; full symbols NPD data).

Similarly to other *RE*FeAsO compounds,[39] for pure LaFeAsO the structural transition is first order, as revealed by the phase coexistence detected between ~ 140 K and 150 K by NPD analysis.[40] This same phenomenon takes place in the $La(Fe_{0.90}Ru_{0.10})AsO$ sample in a similar and quite wider temperature range; in Figure 5, on the left, the peak intensities at the tetragonal 110 and orthorhombic 020 reflections are plotted as a function of temperature. This behaviour is consistent the coexistence between the two polymorphs in this temperature range, a conditions fulfilled by a first-order character of the transformation. Figure 5, on the right, shows the thermal dependence of the ratio between the $\chi^2$ values obtained by applying the *P4/nmm* or the *Cmme* structural model in the refinements of the NPD data; the orthorhombic model clearly fits better the data collected at $T \leq$ 120 K. Between 130 K and 150 K the two models are equivalent, corroborating the phase coexistence scenario. The first order character of the structural transition in this sample is definitively confirmed by the thermal dependence of the structural order parameter, whose analysis and discussion is deferred in §3.4.

On these bases a two-phase structural model foreseeing the coexistence of both polymorphs was tested to fit the data collected between 120 K and 150 K; Figure 5 on the right shows the so obtained dependence of the orthorhombic phase weight percentage with temperature.

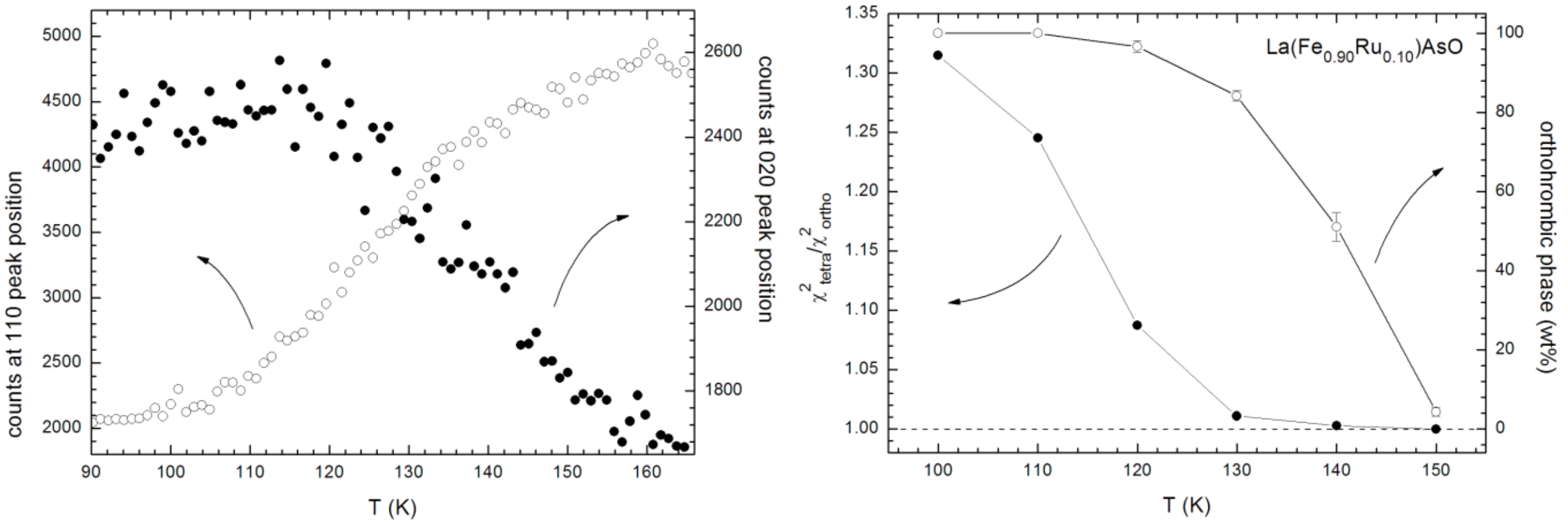

Figure 5: Evidences for a first-order structural transition in La($Fe_{0.90}Ru_{0.10}$)AsO. On the left: Thermal dependence of the peak intensities at the tetragonal 110 and orthorhombic 020 reflections (SPD data). On the right (data obtained after Rietveld refinement of NPD data): on the left scale is plotted the ratio between the $\chi^2$ values obtained using a tetragonal and an orthorhombic structural model (values larger than 1.0 indicate that the orthorhombic model provides a better fit than the tetragonal one); on the right scale is plotted the thermal dependence of the weight percentage for the orthorhombic phase.

Generally speaking a component of the NPD patterns can contain a magnetic scattering contribution if the material undergoes magnetic ordering. At low temperature the ordered magnetic moment value per Fe is lower than 1 $\mu_B$ in pure LaFeAsO;[7] this implies that the magnetic scattering contribution is weak in our NPD patterns and in fact it can be hardly appreciated for the La($Fe_{0.90}Ru_{0.10}$)AsO sample only (D1B data). In this case a faint signal originated by magnetic ordering is observed at $d \sim 4$ Å below $T_m \sim 116$ K, in good agreement with the $T_m$ measured by μSR (119 K). This is about the same temperature as reported for Pr($Fe_{0.90}Ru_{0.10}$)AsO,[25] confirming the experimental observation that the Fe ordering temperature in *RE*FeAsO compounds seems to be not dependent on the rare earth element being magnetic or not.[7] The increase of the Ru content above $x = 0.10$ decreases the ordered magnetic moment below the instrumental detection threshold.

*3.3 Micro-structural analysis*

Lattice micro-strain along a particular crystallographic direction is originated by local variations of the interplanar spacing along it and can be determined by diffraction line profile analysis. Figure 6 shows the Williamson-Hall plot as obtained using the SPD data collected at 300 K for the $La(Fe_{0.90}Ru_{0.10})AsO$ sample, selected as representative; for clarity only the orders of reflections 00*l*, *h*00 and *hh*0 are reported. A strong micro-strain along the 00*l* direction is evident, whereas along both *h*00 and *hh*0 such strain is notably reduced. The Williamson-Hall plots obtained for the other compositions are qualitatively similar; with the increase of the Ru content, the micro-strain increases along each crystallographic direction, up to a maximum value experienced by the $La(Fe_{0.50}Ru_{0.50})AsO$ composition. Further increase of the Ru content leads to a slight decrease of lattice micro-strain (Figure 7).

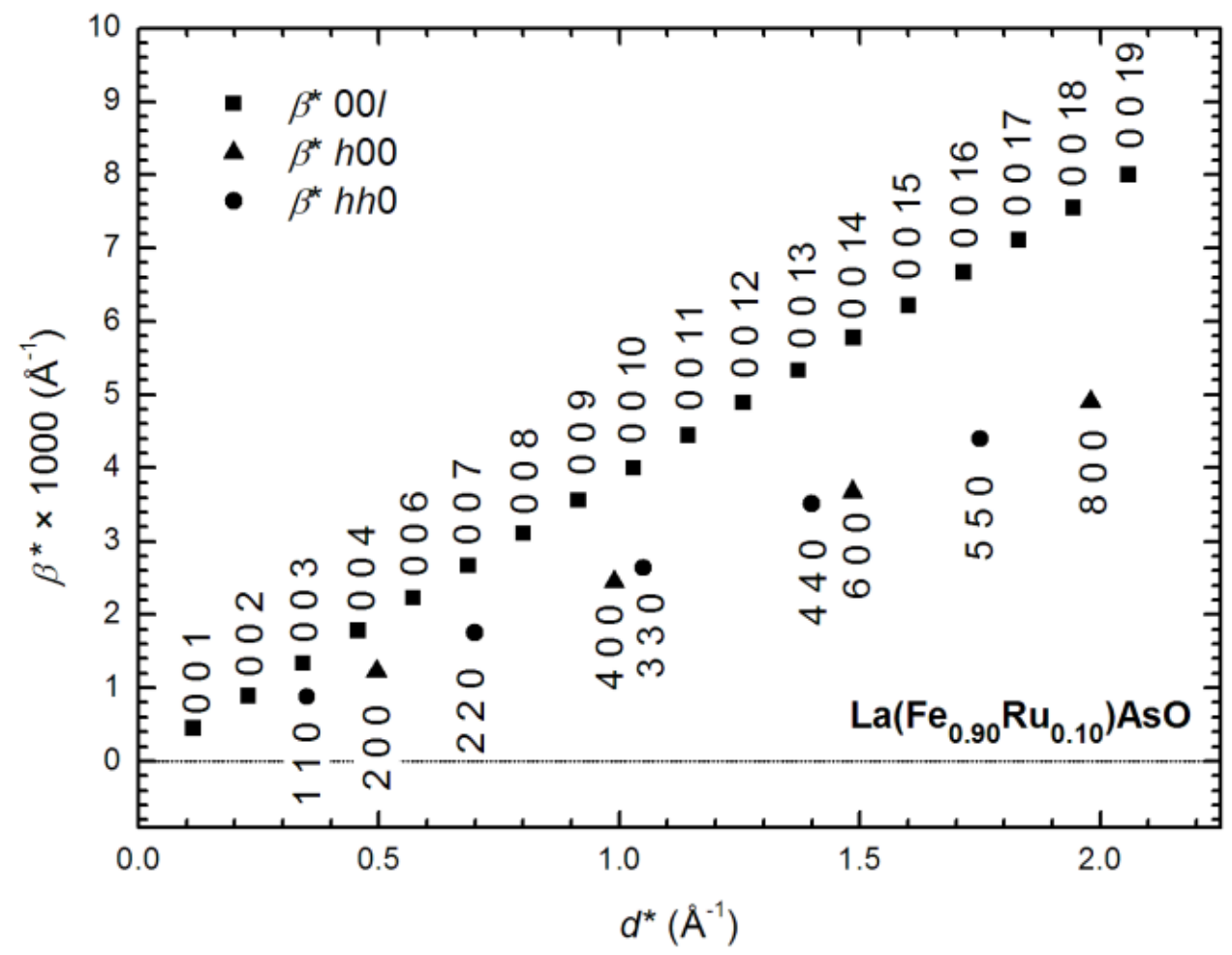

Figure 6: Williamson-Hall plot obtained for the $La(Fe_{0.90}Ru_{0.10})AsO$ sample (from SPD collected at 300 K); for clarity only selected orders of reflections are reported; represents the integral breadth of the diffraction peak.

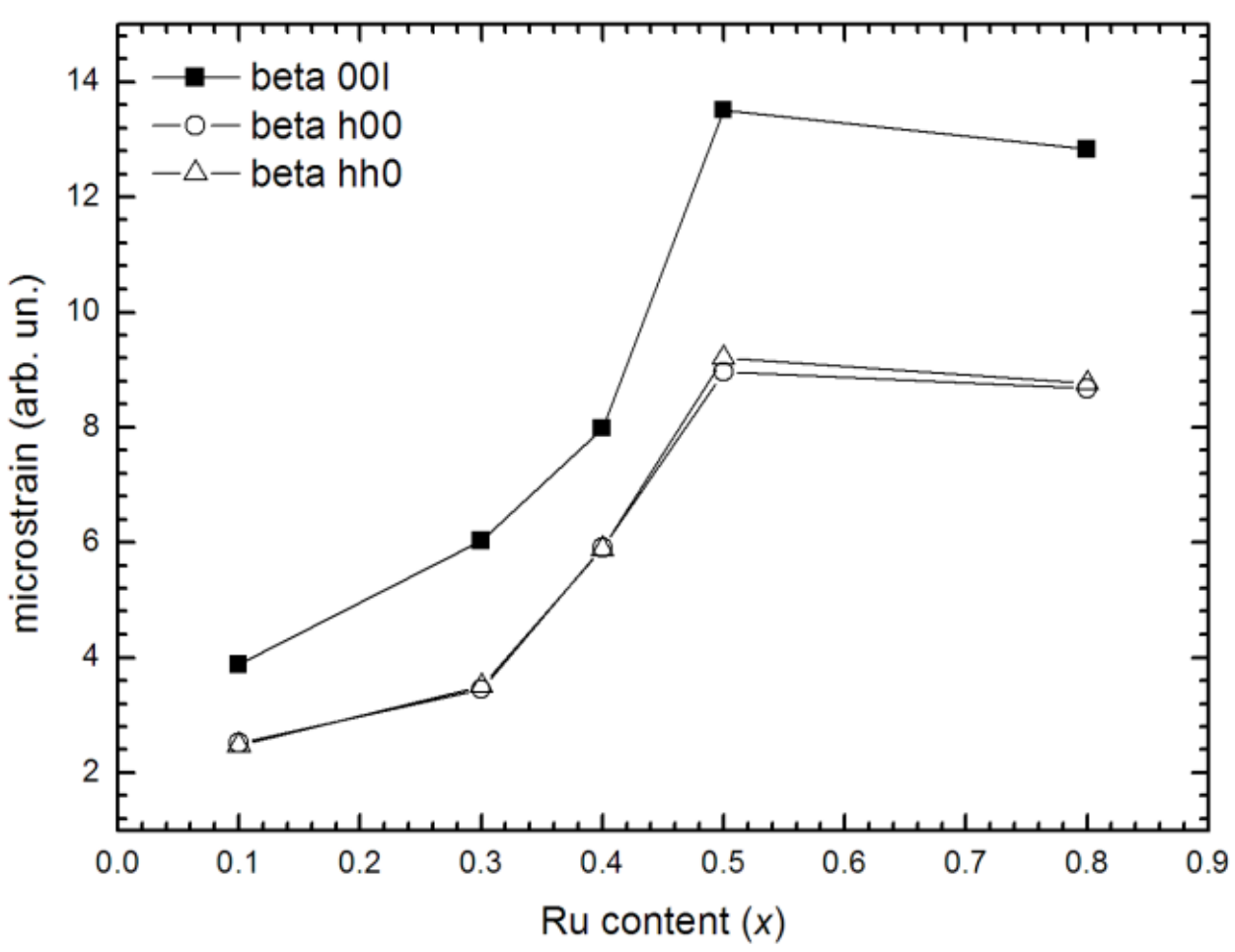

Figure 7: Qualitative evolution of the strain along the three main 00*l*, *h*00 and *hh*0 crystallographic directions as a function of Ru content (from SPD collected at 300 K; lines are guides to the eye).

This behaviour could be related to the mismatch of the ionic radii of Fe and Ru, but even the occurrence of chemical correlations, e.g. a not completely random distribution of elements within the transition metal sub-lattice driven by chemical properties, could be hypothesized. A not homogenous distribution of Ru should determine variations of lattice parameters among crystallites characterized by slightly different composition; this phenomenon causes line broadening on account of a superposition of sub-line profiles with different positions. In such a case, in the tetragonal system, peak broadening depends on the square cosine of the angle $\varphi$ between the diffraction vector and the *c* direction;[41] hence an increase of the peak width as the 00*l* directions is approached can be indicative of a not homogenous distribution of Ru in the structure. This is in agreement with [75]As nuclear quadrupole resonance measurements[24] performed on $Sm(Fe_{1-x}Ru_x)As(O_{0.85}F_{0.15})$, as well as on the same set of samples here investigated,[42] which suggest a tendency towards clustering, that is a non random distribution of Ru. It is worth to note that such a behaviour is not related to an inadequate synthesis process, but instead to chemical relationships favouring some type of positive correlation among the same chemical species at the transition metal sub-lattice. In this context this hypothesis complies to a phase separation scenario proposed for the $Sm(Fe_{1-x}Ru_x)As(O_{0.85}F_{0.15})$ system, where the force constants of Fe-As and Ru-As bond lengths are both unaffected by the degree of substitution,[22,43] even though a simple not homogenous distribution of Ru, with local enrichment, better describes the actual scenario.

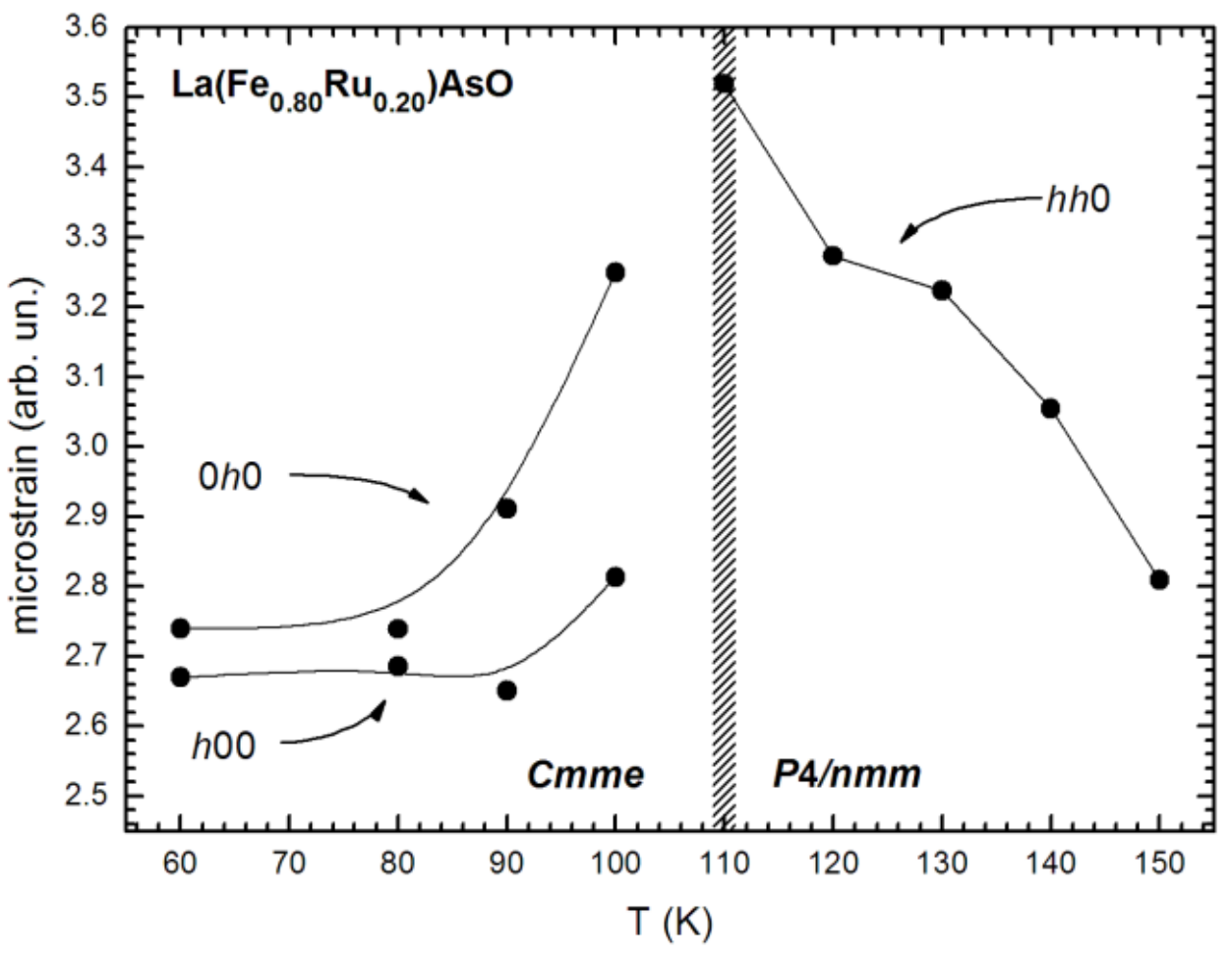

Figure 8: Thermal evolution of the lattice micro-strain along *hh*0 in the tetragonal field and along *h*00 and 0*h*0 in the orthorhombic field of $La(Fe_{0.80}Ru_{0.20})AsO$ (NPD data; lines are guides to the eye).

As the temperature is decreased a different type of micro-strain arises, related to a structural evolution of the crystal lattice as the symmetry breaking is approached. The thermal dependence of micro-strain for samples with $x \leq 0.30$ (Figure 8) is substantially similar to that previously reported for isostructural compounds:[19] a strong increase of the lattice micro-strain along (*hh*0) is detected in the tetragonal structure on cooling, reaching its maximum just above the critical temperature. With the symmetry breaking and the development of the orthorhombic structure, micro-strain is rapidly suppressed. In a previous investigation we have shown that this kind of lattice micro-strains, when detected in the tetragonal phase, are the signature of a tendency of Fe-Fe bond lengths to branch.[19]

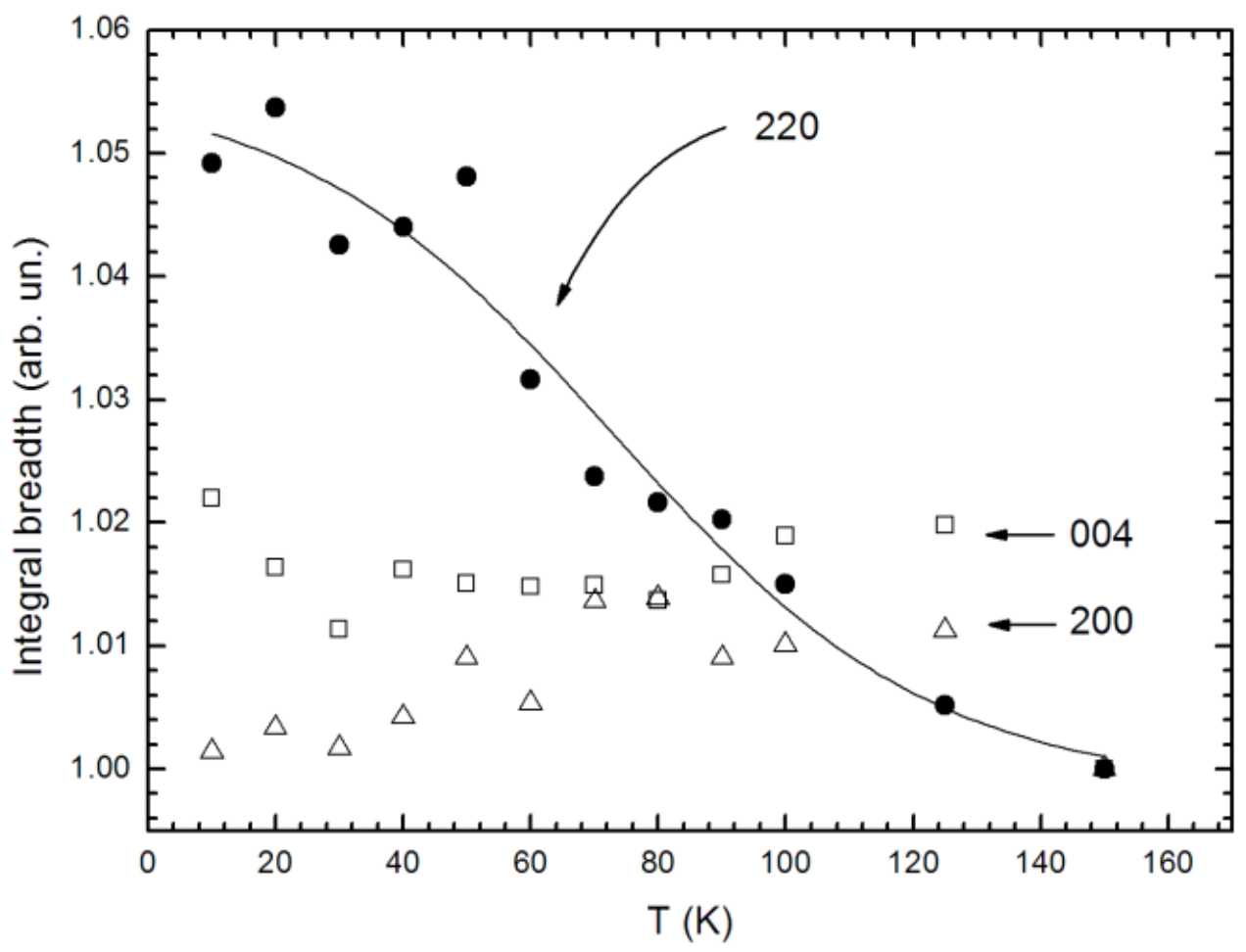

Figure 9: Thermal evolution of the integral breadth due lattice micro-strain affecting three main diffraction peaks in $La(Fe_{0.60}Ru_{0.40})AsO$; (NPD data normalized for the value @150 K; line is a guide to the eye).

In the $La(Fe_{0.60}Ru_{0.40})AsO$ sample, where symmetry breaking is suppressed, a detectable increase of the lattice micro-strain along (*hh*0) can also be observed below ~ 100 K (in Figure 9 this micro-strain is measured as the integral breadth of the diffraction line). This phenomenon indicates that the tendency of Fe-Fe bond lengths to branch is still active on a local scale in this sample, even though the lattice symmetry as a whole remains tetragonal.

*3.4 Spontaneous strain*

The spontaneous strain was investigated by analyzing the thermal dependence of the symmetry breaking and non-symmetry breaking components;[44,45] at this scope the lattice parameters of the tetragonal phase were extrapolated from the measured values above the structural transition temperature.

Spontaneous strain is the lattice distortion solely induced by the structural transition; by applying the general equations for calculating the components of the spontaneous strain ,[44,45] the strain behaviour for the samples undergoing the structural transitions ($x$ = 0.10, 0.20 and 0.30) has been studied, using the NPD data. The proper symmetry-breaking component of the spontaneous strain for the $P4/nmm \rightarrow Cmme$ structural transition is ,[45,46] having the same symmetry of the order parameter Q ( $\propto$ Q); the component is not-symmetry breaking, whereas and are both constituted by a symmetry breaking component plus a non-symmetry breaking one.

In order to take into account the quantum saturation of the phonon modes taking place at low temperature, cell parameters for the tetragonal phase have been extrapolated at low temperature in the orthorhombic field, by fitting the measured cell edges $a$ and $c$ above the structural transition with modified Eq. (1) and (2). Hence in this case was fixed to the value obtained from the cell volume fitting, whereas , and (or ) were the free parameters.

The temperature evolution of Q (obtained by normalizing the values the symmetry breaking strain component ) for La($Fe_{0.90}Ru_{0.10}$)AsO is reported in Figure 10: at $T_s$ the phase transition is stepwise, then the order parameter Q increases down to ~ 100 K, below which its values levels off. The symmetry breaking components = (not shown) exhibit a similar thermal dependence, even though does not vary linearly with , implying a different strength of coupling between Q and or (Figure 11). The variation of the order parameter Q with temperature displayed in Figure 10 is typical of a structural transition with a first order character,[44] in agreement with the conclusions drawn in § 3.2.

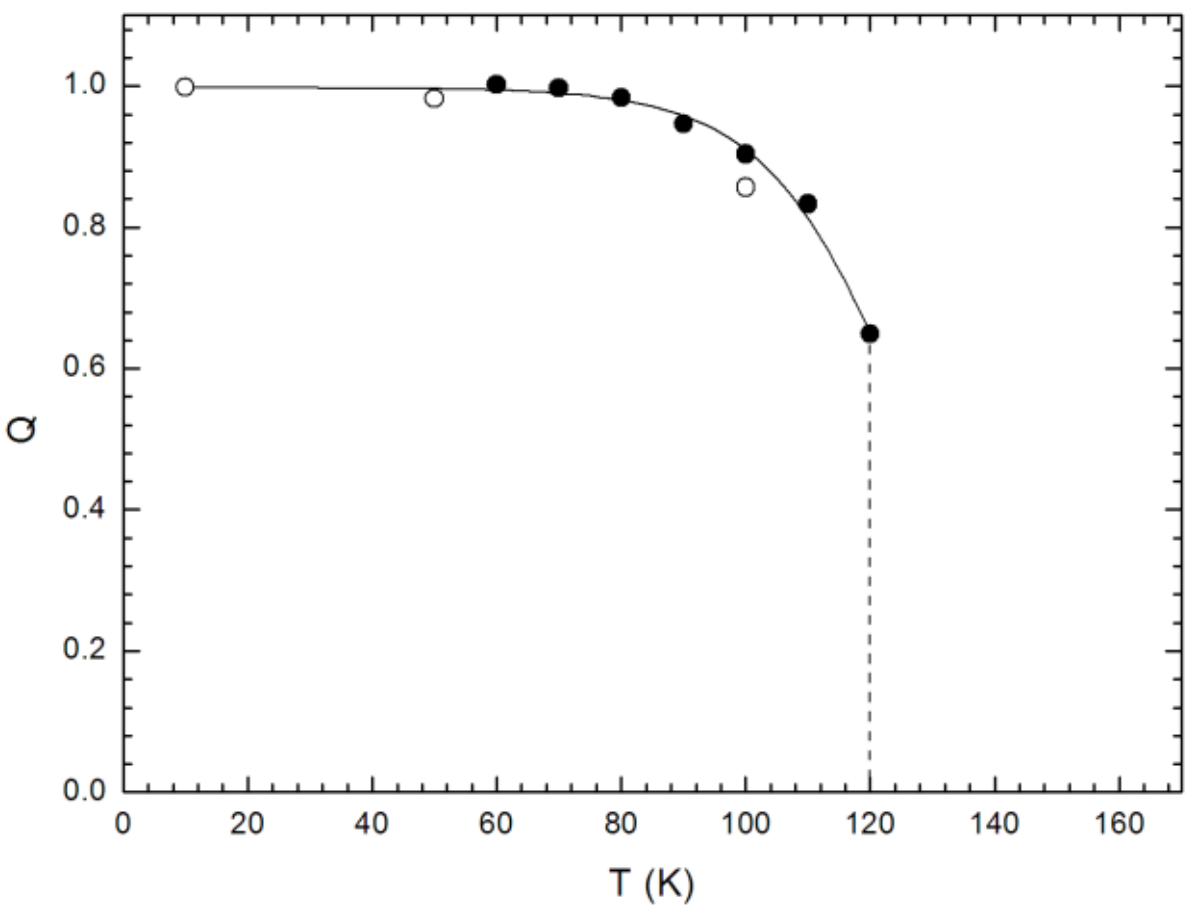

Figure 10: Variation of the order parameter through the structural transition in La($Fe_{0.90}Ru_{0.10}$)AsO, consistent with a first order character of the structural transition (open symbols: SPD data; full symbols NPD data); lines are guides to the eye.

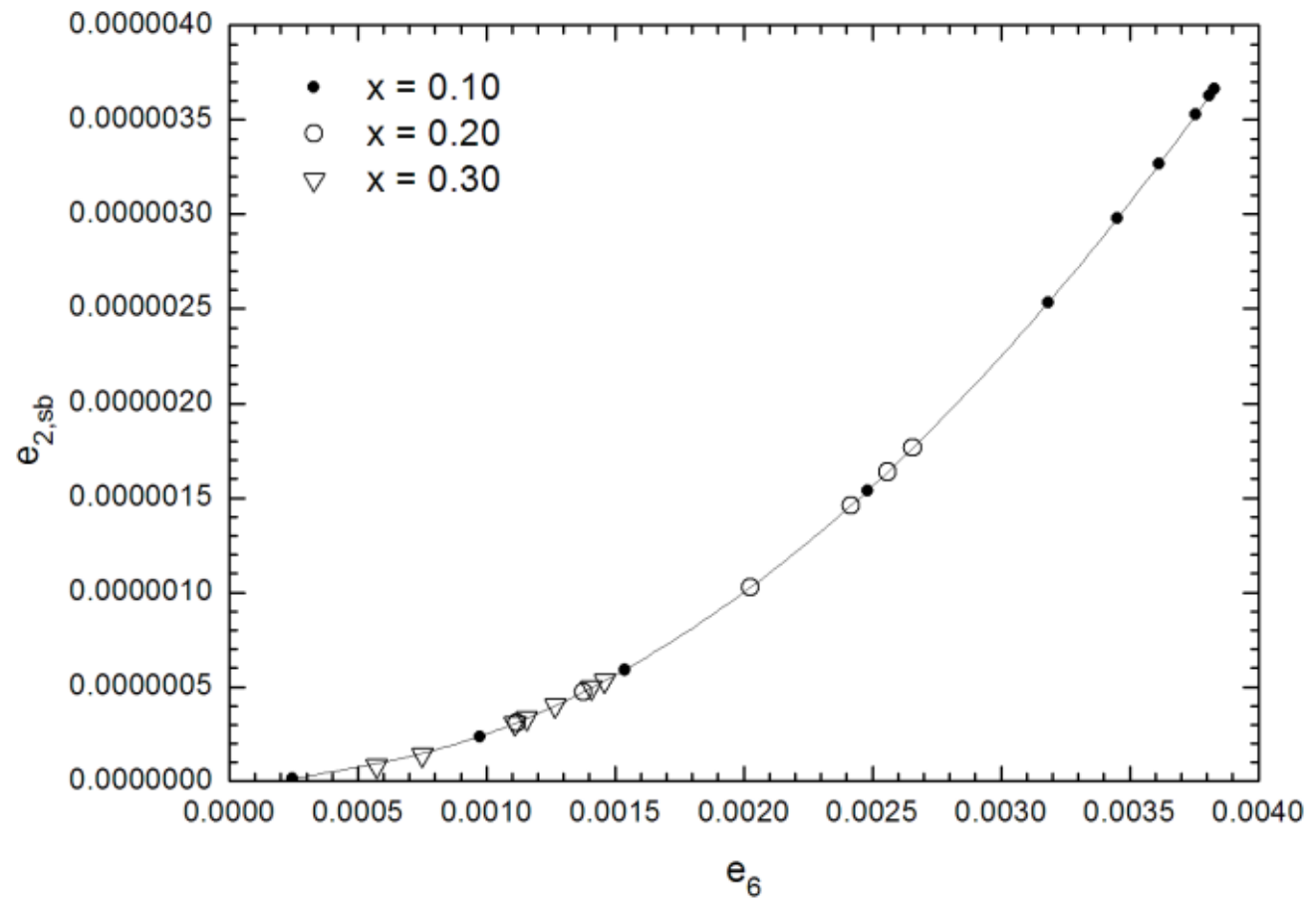

Figure 11: vs in the La($Fe_{1-x}Ru_x$)AsO samples undergoing the structural transition; the line is a guide to the eye.

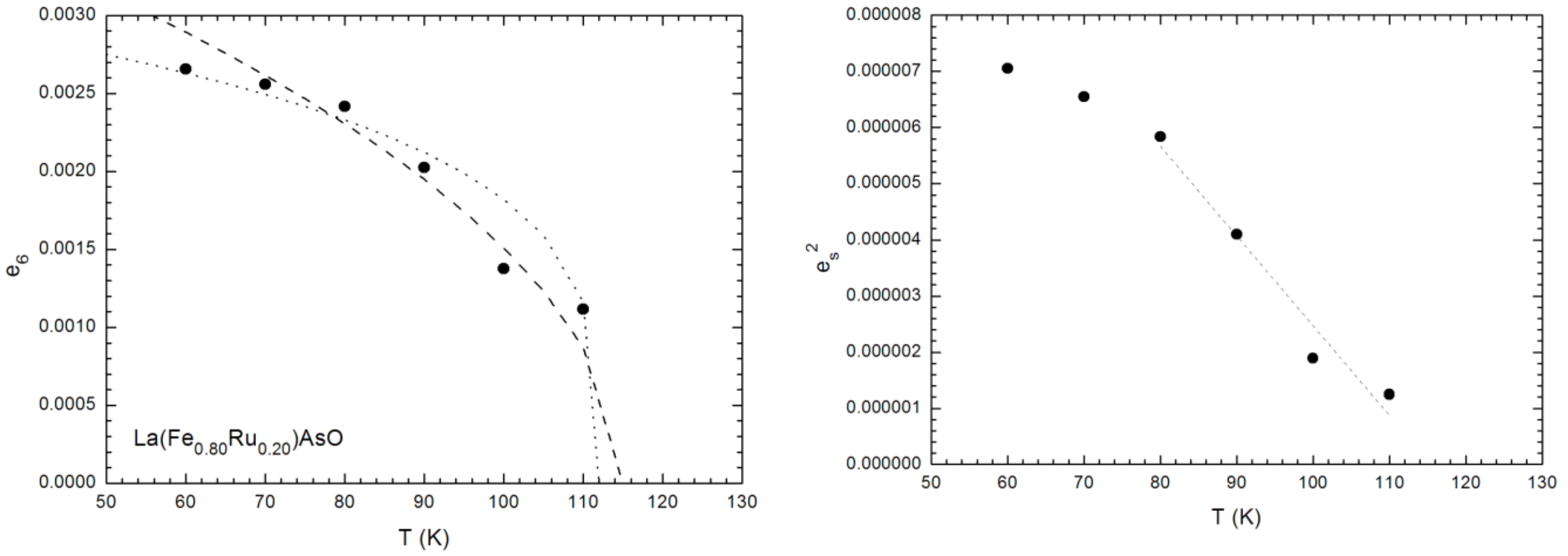

Figure 12: On the left: Variation of the symmetry breaking strain component through the structural transition in La($Fe_{0.80}Ru_{0.20}$)AsO; the dotted and dashed lines correspond to the predicted behaviour of a tricritical and second order transition, respectively. On the right: the square of the spontaneous strain is linear with temperature between ~ 80 and ~ 110 K.

In La($Fe_{0.80}Ru_{0.20}$)AsO the onset of the structural transition is at ~ 110 K; due to the reduction of the orthorhombic distortion a strong peak superposition occurs and hence a reliable conclusion about the possible phase coexistence cannot be gained, neither by NPD nor by SPD data. The evolution of the symmetry-breaking component with temperature is reported in Figure 12, on the left. The contribution of the other strain components to the spontaneous strain is negligible and hence . The almost linear evolution with *T* of over a temperature range between 80 and 110 K suggests a tricritical or second order behaviour (Figure 12, on the right).[44] The Landau theory predicts and for a second order and a tricritical phase transition, respectively,[44,45] but the paucity of the data does not allow for any firm conclusion (Figure 12, on the left).

The La($Fe_{0.70}Ru_{0.30}$)AsO sample displays a further evolution of the transition behaviour. Figure 13, on the left, shows the thermal dependence of the symmetry-breaking component ; this data can be satisfactorily fitted with the behaviour for a second order transition in the Landau theory. In addition is linear with temperature over the entire temperature interval, as expected for a second order structural transition (Figure 13, on the right),[44] and its linear fit extrapolates to zero at ~ 82 K, in perfect agreement with the $T_s$ value obtained by analyzing the 110 peak broadening.

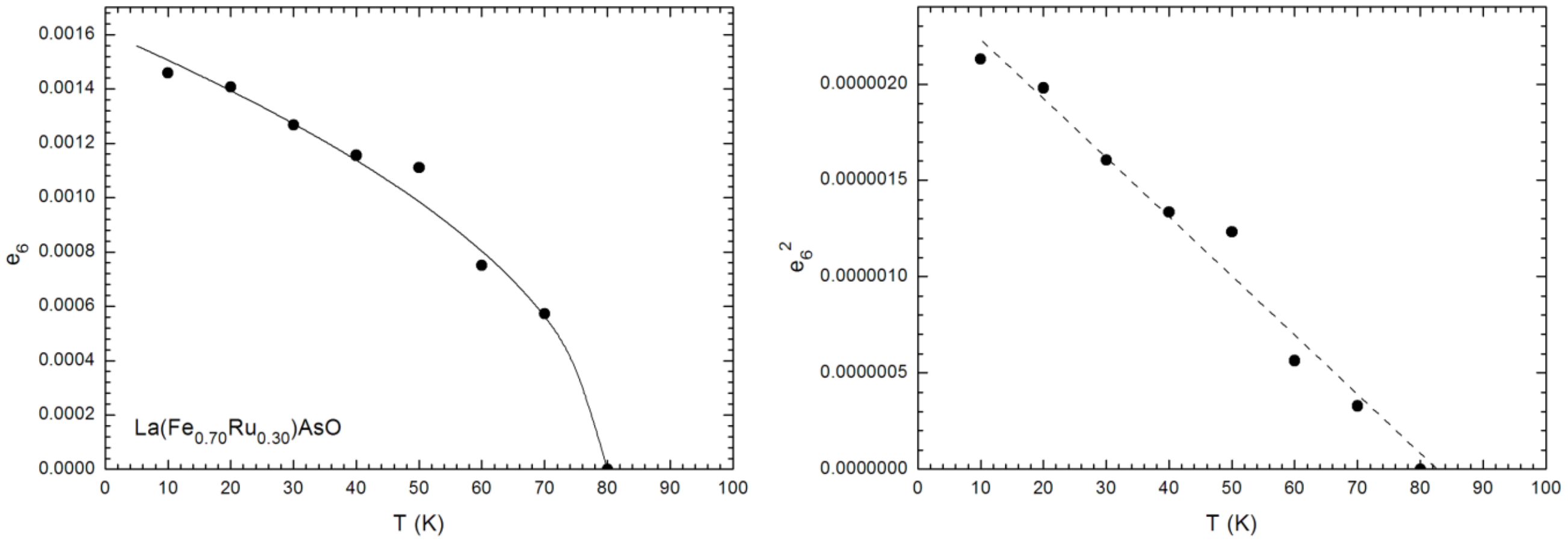

Figure 13: On the left: Variation of the symmetry breaking strain component in La($Fe_{0.70}Ru_{0.30}$)AsO; the line corresponds to the predicted behaviour of a second order transition. On the right: Linear variation with temperature of the square value of $e_6$.

Finally it is worth to note that the strength of coupling between the order parameter Q and symmetry-breaking components and does not change with Ru content, as can be inferred from Figure 11; the data plotted in this figure were collected at different temperatures, but their position depends uniquely on the degree of orthorhombic distortion, since:

(3)

when ~ 90°, as in our case ( , , are the cell parameters of the orthorhombic low-symmetry phase; is defined as in figure 9 of ref. [46]). This figure clearly demonstrates that the orthorhombic distortion develops in the same way in the analyzed samples, independent of the composition and the temperature at which it occurs, however, its evolution is progressively hindered with the increase of Ru content.

As a conclusion, from these analyses it can be argued that as the amount of Ru content increases the character of the tetragonal to orthorhombic structural transition changes from first to second order.

**4. Discussion**

By combining the structural data of the present work with the magnetic one obtained on the same sample series by μSR analysis,[29] a tentative phase diagram for the La(Fe,Ru)AsO system has been drawn in Figure 14, assessed in order to comply to thermodynamic rules.

For pure LaFeAsO the reported onset of the structural transition occurs around 155 – 160 K,[40,47] but the tetragonal structure was found to coexist with the low symmetry phase down to ~ 140 K.[40] The phase rule imposes that when the two polymorphs of pure LaFeAsO are coexisting, the variance of the system is 0 and as a consequence the tetragonal phase found below 155 – 160 K is in a metastable state. After Ru-substitution the system gains one more degree of freedom and in this case thermodynamic predicts that a 2-phase field should separate into two single-phase fields when a first-order structural transition takes place; conversely, two single-phase fields can be in contact when the character of the transition is second-order. The character of the structural transition in $La(Fe_{1-x}Ru_x)AsO$ is first order for $x \leq 0.10$, since experimentally the coexistence of the tetragonal and the orthorhombic phase has been observed between ~ 160 K and ~ 120 K; hence a 2-phase field separates the stability fields of the tetragonal and orthorhombic phases within this compositional range. As the character of the transition becomes second-order the 2-phase field is suppressed and a single boundary separates the tetragonal phase field from the orthorhombic one; for $x = 0.10$ the 2-phase field is evaluated as reported in § 3.4.

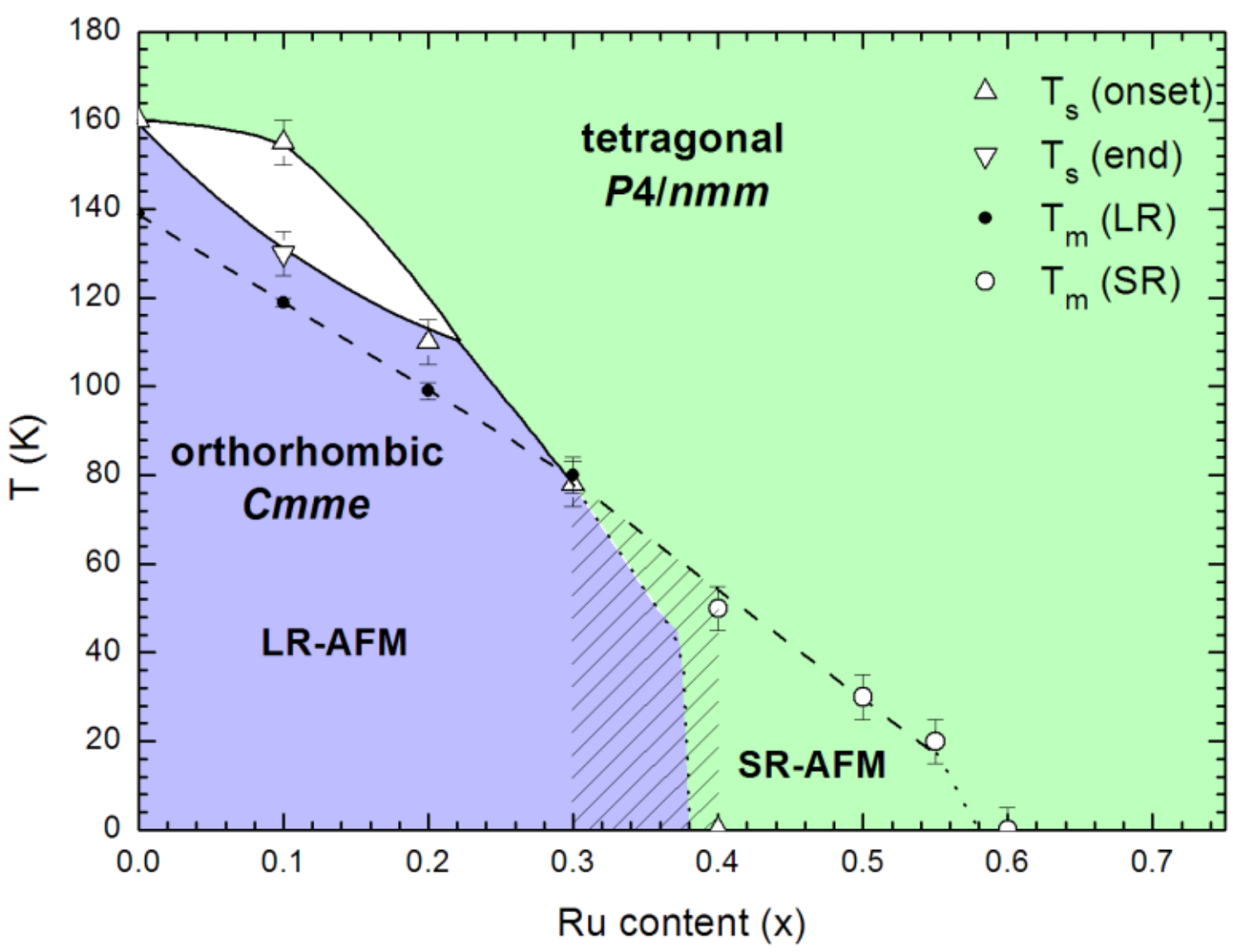

Figure 14: The phase diagram of the La(Fe,Ru)AsO system. The dashed line marks the stability field of the magnetic ordering; dotted lines stand for estimated phase boundaries; the hatched area indicates the crossover from a long- to a short-range magnetic ordering.

The only slight decrease of the onset temperature of the structural transition observed for $x \leq 0.10$, indicates that the strain fields around each solute Ru atom do not overlap at this composition. In fact the local deformation induced by the dissolved atom can be absorbed by the structure by distortions within the tetrahedra; the length scale of the relaxation can be calculated from the extent of the plateau at the end of the phase diagram ($x$ = 0.10), corresponding to a ratio of 1:5 unit cells containing a Ru atom (each unit cell contains 2 unit formula). This composition can be considered the limit above which the strain fields around each Ru atom begin to overlap, since $T_s$ undergoes a net decrease for higher Ru content. For this composition a uniform distribution of Fe and Ru atoms at the transition metal site implies an average Ru-Ru separation of 2 unit cells and hence a distorted region around each Ru atom of about 2.5 cells across (~ 10 Å).

As the Ru content increases strain fields overlap and the structural transition progressively decreases in temperature as well as in amplitude; symmetry breaking is totally suppressed for $0.30 < x < 0.40$. Within this compositional range the evolution of the phase boundary can be roughly estimated taking into account quantum mechanical effects. In fact at low temperature quantum fluctuations modify the behaviour of the structural transition, enhancing the stability of the high-temperature phase.[48] In particular below ~ 0.25 the temperature dependence of the order parameter Q is dominated by quantum mechanical effects and is independent on temperature; in our case this occurs below ~ 50 K. This evaluation is corroborated by the data reported in Figure 10, on the left, showing that the order parameter Q saturates below ~ 60 K. On these bases the phase boundary for $0.30 < x < 0.40$ was traced in Figure 14 (dotted line).

Regarding the comparison between the structural and magnetic properties, it is important to emphasize here that clear oscillations of the time dependence of the μSR asymmetries, for $T \ll T_m$, are present only for $x \leq 0.3$, whereas they are absent in samples with $x \geq 0.4$ (see Fig. 2 of ref. [29]). In fact in the latter the oscillations are markedly damped because of the increase in the width of the local field distribution at the muon site. This behaviour is due to the fast loss of the coherence of the muon spin precessions around the local field,[52] and indicates a crossover from a long-range (LR) to a short-range (SR) magnetic order for $x$~0.35.

In Figure 14, for $x \leq 0.30$, the magnetic transition temperatures, $T_m$, can be directly determined from the evolution of the mean magnetic order parameter <S(T)> as a function of temperature, which was obtained from the muon spin precession around the local muon field reported in Figure 4 of Ref. [29]. The magnetic order parameter <S(T)> has been successfully fit to the phenomenological function $\langle S(T)\rangle = S(0)\ [(1-T/T_m)^{2.4}]^{0.24}$, which is found to hold generally for *RE*FeAsO compounds.[49,50] For each composition the fit has been applied only to those data points which correspond to temperatures where the sample volume is almost fully magnetic, in order to consider the magnetic response coming from the whole sample. Unfortunately, this procedure cannot be applied for those samples with $x > 0.30$ which do not display coherent precessions of the muon spin due to the lack of long range magnetism and the behavior of S(T) cannot be inspected.[29] For these samples $T_m$ is taken as the temperature at which the magnetic volume fraction is 80%, which incidentally has been found for $x \leq 0.30$ to be the average value of the magnetic volume fraction at $T_m$, as determined by the fit of <S(T)>. Under these assumptions the magnetic properties exhibit a homogeneous behaviour; the magnetic transition temperature $T_m$ almost linearly decreases up to $x$ = 0.50; with a further increase of the Ru content the decrease of $T_m$ is enhanced down to its suppression for $x \sim 0.60$, similarly to what has been observed in the Ce(Fe,Ru)AsO system.[26] In the Pr(Fe,Ru)AsO phase diagram[25] the magnetic ordering of Fe spins is reported to extend up to $x$ = 0.10 only, however, one has to note that this result is based on NPD data only; this technique is much less sensitive than μSR when the correlation length and magnetic scattering intensity is reduced by the Ru doping. Noteworthy the same authors infer the suppression of magnetism by transport properties in the same series of materials at $x = 0.67$ in a previous work.[20] Also in our NPD data on La(Fe,Ru)AsO the weak characteristic magnetic reflection peak at $d \sim 4$ Å is perceptible only up to $x = 0.10$. However the muons, in virtue of the short range character of the dipolar interaction between the muon ½ spin and the Fe magnetic moments, are local probes very sensitive to the presence of magnetic ordering even when the magnetic coherence length $\xi$ becomes of the order of few nanometers. In particular when $\xi$ becomes smaller than ~10 lattice spacing the muon precessions around the magnetic local fields become incoherent and a strong μSR signal decay appears instead of coherent oscillations.[51] As reported above, in our La(Fe,Ru)AsO samples the lack of coherent oscillations occurs for $x \sim 0.35$,[29] indicating that the system is driven from a long-range, with ξ » 1 nm, to a short-range, with ξ~1 nm, AFM ordering. This is in the same compositional range where the orthorhombic structure is suppressed. The LR-AFM ordering always takes place within the orthorhombic structure, suggesting that it can only develop after the symmetry decrease. Conversely SR-AFM ordering takes place in the tetragonal phase; the micro-structural analysis of the sample with $x = 0.40$ reveals that magnetism occurs within the temperature range where lattice micro-strain appears, that is when Fe-Fe bond lengths tend to branch even though only on a local scale. This result suggests that even in this case magnetism is driven by symmetry breaking, although confined on a local scale.

The origin of this SR magnetism seems different from that observed in the CeFeAs($O_{1-x}F_x$) system, where the LR magnetic order observed in CeFeAsO gradually turns into a SR order with the increase of the electron doping.[52] Conversely in the La(Fe,Ru)AsO system the interactions among Fe ions are progressively prevented by the progressive increase of non-magnetic Ru atoms.[23] In this case magnetic interactions can develop in a LR structure when the concentration of Fe ions is sufficiently high, but with the increase of substitution they remain progressively confined at a shorter and shorter scale. It is interesting to note that both the long range magnetism and the orthorhombic structure are suppressed around $x = 0.4$ which is the percolation threshold of the Ru enriched cluster expected in a diluted square lattice. The presence of SR magnetism above this threshold implies that magnetically ordered Fe islands of nm size may survive interspersed within a magnetically disordered background in the (Fe,Ru)As layer. Nevertheless, the lattice micro-strain detected also at $x = 0.4$ at the same temperature where the SR magnetism develops, is strongly indicative that the symmetry breaking, at least on a local scale, is always concurrent with magnetism. Again one has to conclude that the magnetism is always driven by structural symmetry breaking, even though confined here to a local scale.

In order to highlight the role of the structural disorder, the dependence of the magnetic transition temperature $T_m$ on the variance in the distribution of the transition metal ionic radii was investigated; in fact is a measure of the structural disorder induced by the pseudo-random distribution of cations with different radii at the same atomic site.[53] Unfortunately, in our case, only the ionic size of $Fe^{2+}$ in tetrahedral coordination is available in literature (0.63 Å),[54] but not that of $Ru^{2+}$. In order to roughly estimate this value, the volumes of the square prism inscribing the [$FeAs_4$] and [$RuAs_4$] tetrahedra in LaFeAsO and LaRuAsO were calculated as:

$$- \quad - \quad — \qquad (4)$$

Where $a$ and $c$ are the cell edges of the end members calculated by the linear relationships given in §3.2 and $z$ is the coordinate of As for pure LaFeAsO and LaRuAsO, estimated by noting that it roughly linearly increases with composition (Table 1). Finally it was assumed that the ratio between the so-calculated volumes is proportional to the ratio between the ionic radii of $Fe^{2+}$ and $Ru^{2+}$. As a result the ionic radius of $Ru^{2+}$ is estimated to be ~ 0.644 Å and the variation of $T_m$ as a function of is reported in Figure 15, evidencing a decrease of $T_m$ not linear in . This behaviour indicates that the magnetic transition temperature is not solely dominated by disorder, but spin dilution accompanying Ru substitution is effective also in the Fe-rich side of this system, in agreement with previous findings.[29] In particular for $x \leq 0.30$ frustration plays only a secondary, but detectable role; with the increase of the Ru content the percolation threshold for the $J_1$ - $J_2$ model on a square lattice (corresponding to $x \sim 0.60$) is approached[29] and a dramatic decrease of $T_m$ is observed.

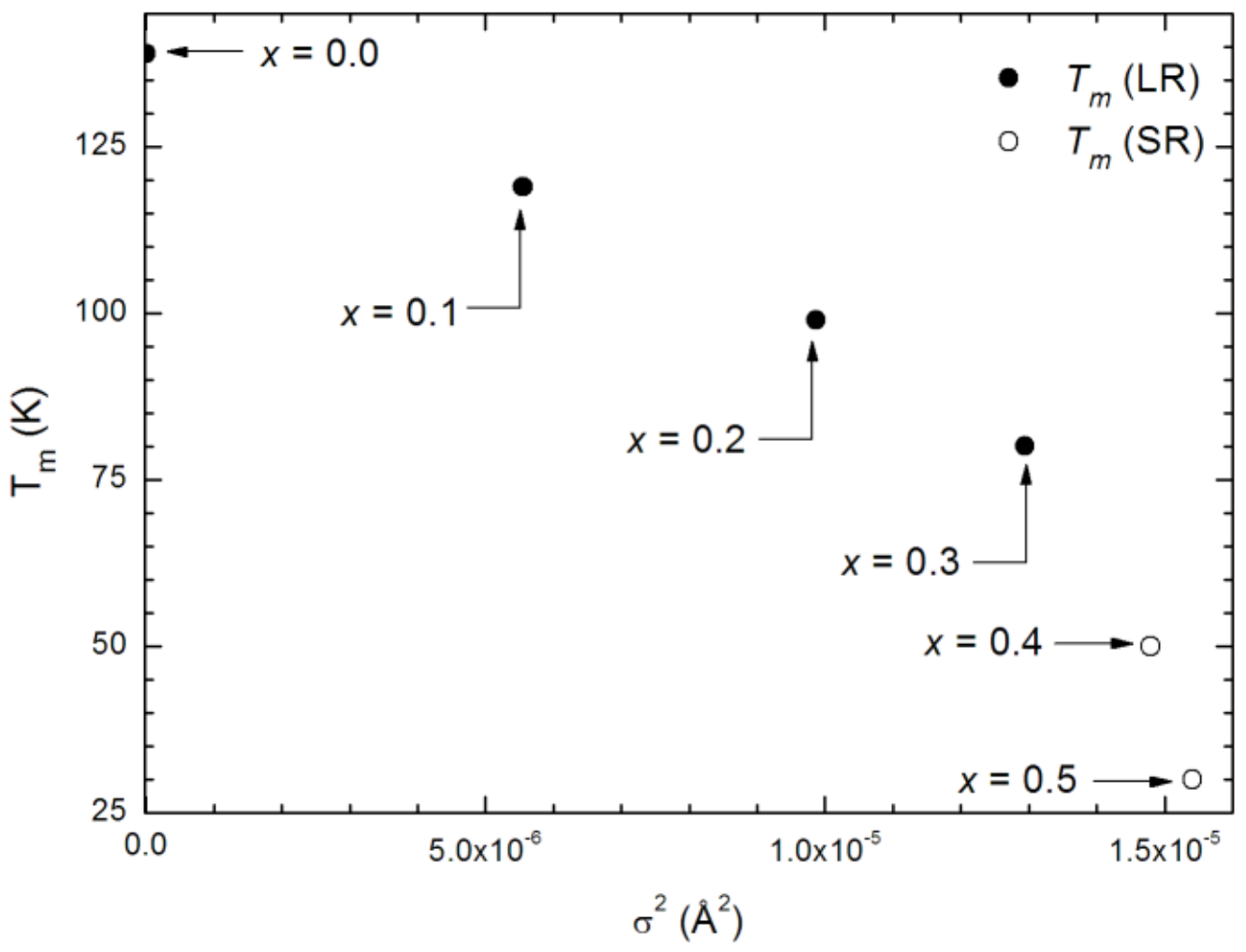

Figure 15: Variation of the magnetic transition temperature $T_m$ with cationic variance in the La(Fe,Ru)AsO series.

## 5. Conclusions

In conclusion the crystal structure, the micro-structure and the spontaneous strain characterizing selected members of the $La(Fe_{1-x}Ru_x)AsO$ solid solution have been investigated, using high resolution neutron and synchrotron powder diffraction data collected between 10 and 300 K. Lattice parameters follow the Vegard's law with opposite trends and a tetragonal to orthorhombic structural transition takes place on cooling up to $x$ = 0.30. In agreement with NQR and EXAFS measurements, microstructural analysis suggests a scenario where chemical correlations take place at the transition metal sub-lattice, determining a non random distribution of Ru. A notable increase of lattice strain occurs along *hh*0 just above the symmetry breaking, suppressed with the onset of the orthorhombic symmetry. Spontaneous strain analysis indicates a crossover from a first to a second order character of the structural transition with the increase of Ru substitution. No negative thermal expansion occurs in the $La(Fe_{1-x}Ru_x)AsO$ system, differently to what is observed in the homologous $Pr(Fe_{1-x}Ru_x)AsO$ one. As long as the tetragonal-orthorhombic transition is retained the magnetic ordering of the Fe sub-lattice is long-range, but with its suppression magnetism becomes short range ordered and the symmetry breaking gets a local character. Isoelectronic Ru substitution then induces two main effects on the magnetism of the system: 1) disorder at the magnetic sub-lattice; 2) frustration of the Fe magnetic moment ions, effective also in the Fe-rich side of the system. Finally a phase diagram for the La(Fe,Ru)AsO system has been drawn by combining the structural and magnetic data. The results clearly support that magnetic and structural properties are strongly correlated and that magnetism is always driven by symmetry breaking, even when both are confined on a local scale.

**Acknowledgments**

A.M. acknowledges dott. G. Lamura (CNR-SPIN) for discussions. S.S. acknowledges the Fondazione Cariplo (research grant n.2011-0266). This work has been supported by FP7 European projects SUPER-IRON (n°283204).